\documentclass{article}[12]
\usepackage{graphicx} 
 
\begin{document}

\markboth{Danuta Makowiec and Wies{\l}aw Miklaszewski}
{Traffic Flow by Cellular Automata: the Effect of Maximal Car Velocity}

%\catchline{}{}{}{}{}
%%%%%%%%%%%%%%%%%%%%%%%%%%%%%%%%%%%%%%%%%%%%%%%%%%%%%%%%%%%%%%%%%%%%%

\title{TRAFFIC FLOW BY CELLURAR AUTOMATA: THE EFFECT OF MAXIMAL CAR VELOCITY} 
 
\author{DANUTA MAKOWIEC and WIES{\L}AW MIKLASZEWSKI\\
       Institute of Theoretical Physics and Astrophysics,\\
        Gda\'nsk University, \\ 
       ul. Wita Stwosza 57,  80-952 Gda\'nsk, Poland,\\
       {\it e-mail} fizdm@univ.gda.pl} 
 
\maketitle

\begin{abstract} 
Effects of large value assigned to the maximal car velocity on the fundamental diagrams in the Nagel-Schreckenberg model are studied by extended simulations. The function relating  the flow in the congested traffic phase with the car density and deceleration probability is found numerically. Properties of the region of critical changes, so-called jamming transition parameters, are described in details. 
The basic model, modified by the assumption that for each car an individual velocity limit is assigned, is investigated in the aim to find the best supplementary rule allowing the jammed traffic to move with velocity larger than the slowest driving vehicle.  
\end{abstract}
 
 
\section{Introduction} 
In recent years vehicular traffic problems have attracted much attention and a number of cellular automata models describing the traffic flow have been proposed in order to consider the dynamical aspects of the traffic system \cite{Review,Helbing}. Presently, there are two basic cellular automata models that describe single lane traffic flow:  the Nagel-Schreckenberg (NaSch) model \cite{NagelSchrekenberger} and the Fukui-Ishibashi model \cite{FukkuiIshibashi}. Besides the cellular automata models, which are discrete in space and time, several other approaches to the  traffic flow have been proposed recently. Among them there are space-continuous models in discrete time such as model of Krauss {\it et al.} \cite{Krauss,Krauss97} or  models continuous in space and time, e.g., the macroscopic (fluid-dynamical) models \cite{fluid}. 

The cellular automata traffic modeling  enables easy examination of microscopic and macroscopic aspects of the traffic flow and  therefore it is considered as an important tool in traffic engineering   \cite{Review,Helbing}. Though the models are fairly simple, it has been shown that they are able to reproduce  real life traffic phenomena such as spontaneous formation of jams \cite{NSreview}, explain
the impact of global traffic light control strategies \cite{NSlights} on conditions for car accidents \cite{NSaccidents,Moussa03}.

The basic NaSch model parameters are: a maximal allowed vehicle velocity $v_{max}$,  a vehicle density $\rho$, and  a random deceleration probability $p$. The mean velocity $v$ of all vehicles on a lane is  the observable measured. At low density, $\rho << 1$, the vehicle flow is characterized by a linear dependence on the vehicle density.  At high densities the vehicle flow decreases with increasing density and vanishes for the  density $\rho=1$.  The vehicles are said to be in the congested state ---  so-called stop-and-go waves dominate in the dynamics of the system. Thus, with increasing car density the traffic state changes from the free flow to the congested traffic. 

The limit $v_{max}=\infty$ has been introduced by Sasvari and Kertesz \cite{SasvariKertesz}. They have found that the fundamental diagram has a form quite different from that for a finite $v_{max}$. The flow does not vanish in the limit $\rho\rightarrow 0$ since even a single car produces a finite value of the flow. Due to the hindrance effect of other cars the flow is a monotonically decreasing function of the density. Hence the jamming transition disappears. 

In the following paper we investigate vanishing of the jamming transition. 
In particular, we study the influence of the general car velocity limit on the traffic features in order to identify the model parameters for which this transition disappears.

Within our model several specific real life observations have been attempted to reproduced. Let us mention the German highway rules \cite{Ebersbach01}, different rules for overtaking, see  \cite{Review} for the such rules review,  or even the individual driver reaction to traffic \cite{Moussa04}. In this context we decided to consider a very Polish rule --- each driver has its own velocity limit. One can give many reasons supporting our assumption: high diversity of vehicles, namely, from sport cars to scrap cars, or the common  driver habit of breaking the speed limit.  However, if at random we assign to each driver his/her own maximal driving speed then the stationary state is determined  by the slowest vehicles.  Therefore, again following hints from the Polish roads, the slowest driver is forced to change his/her behavior. By simulation, we observe the self-organization of the traffic when the maximal velocity of the slowest moving vehicle is changed.

\section{Standard NaSch model }

The probabilistic cellular automata model of a traffic represents a lane as  one-dimensional lattice of $L$ cells.  Each cell is either  occupied by one of $N$ cars or is empty. A car can move with the velocity
determined by integer values bounded by a speed limit $v_{max}$.  At time $t$  a car is identified by a cell number $x_i^{(t)}$ which the car occupies and the velocity $v_i^{(t)}$ with which the car go. The number of not occupied cells before the car is denoted by $g_i^{(t)}=x_{i+1}^{(t)}-x_i^{(t)}-1$ and is usually called the gap. The cars move along the lane following rules related with the driving habits:

\begin{itemize}
\item[A] Since every driver tends to drive with the maximal allowed velocity the car velocity is incremented in every time step ({\bf acceleration}) 
\begin{equation}
	v_i^{(t+1)}=\min(v_i^{(t)}+1,v_{max}).\label{acc}
\end{equation}
\item[B] If there is a risk of collision with the preceding car, the driver decreases the velocity to avoid the crash ({\bf braking})  
\begin{equation}
	v_i^{(t+1)}=\min(v_i^{(t)},g_i^{(t)}). \label{bra}
\end{equation}
\item[C]In order to include the unpredictable factors which influence reactions of the driver the car velocity is decreased with a small probability $p$ ({\bf randomization})
\begin{equation}
\mbox{if random} < p  \mbox{ then }	v_i^{(t+1)}=\max(v_i^{(t)}-1,0).\label{ran}
\end{equation}
\end{itemize}

Finally, the position of the car is updated ({\bf movement})
\begin{equation}
x_i^{(t+1)}=x_i^{(t)}+v_i^{(t+1)}. \label{mov}
\end{equation}

\begin{figure}[!ht] 
\centerline{\includegraphics[width=0.7\textwidth]{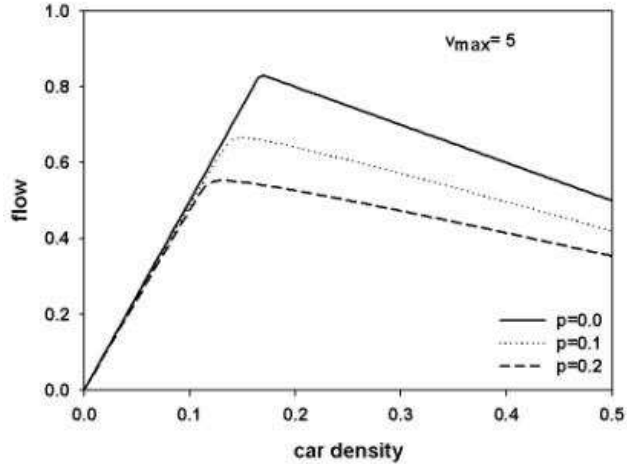}} 
 \caption{The fundamental diagram of the NaSch model for $v_{max}=5$ and different                        random deceleration probabilities $p$.} 
 \label{fig:1} 
\end{figure} 

At each  discrete time step $t \rightarrow t+1$ the positions and velocities of all cars are adjusted in parallel, first, the velocities are obtained by the rules (\ref{acc}-\ref{ran}), then, the positions of the cars are determined according to (\ref{mov}). The periodic boundary condition is applied what effects in that the number of cars is conserved.  

Stationary states of cellular automata with the number conserving rule have in most cases a very simple structure which is reached after $L$ - the lattice size, time steps \cite{BoccaraFuks}. If $p=0$, since the evolution is deterministic, then the stationary state in the NaSch model at low car density is perfectly tiled by the two following patterns
\begin{equation}
(e)\quad \mbox{and} \quad (\underbrace{e,e,\dots,e}_{v_{max}}, v_{max})
\label{free_patterns}
\end{equation}
where $e$ means an empty cell and $v_{max}$ denotes a cell occupied by a car with the actual  velocity $v_{max}$. It is easy to check that this perfect tiling is possible only if the car density $\rho$ satisfies the relation:
\begin{equation}
\rho <\frac{1}{v_{max} +1} = \rho_{crit}
\label{max}
\end{equation}
If the car density $\rho >\rho_{crit} $ then all intermediate to (\ref{free_patterns}) lattice patterns  appear:
\begin{equation}
(0), (e,1),  (e,e,2), \dots, (\underbrace{e,e,\dots,e}_{v_{max} -1 }, v_{max}-1)
\label{congested_patterns}
\end{equation}
If the random deceleration parameter $p>0$ then all patterns from the list (\ref{congested_patterns}) can also be found for $\rho <\rho_{crit} $. The mean velocity is $<v>=v_{max}-p$.  Therefore, the parameter $p$ can be considered as the symmetry breaking field and for the order parameter related to this field $m=v_{max} -<v>$, the thermodynamics characteristics (critical exponents, scaling relations) of the second order phase transition can be estimated \cite{BoccaraFuks}. The similar results can be also found by the mean field approach \cite{Schadschneider}.

In order to use the empirical results for a theoretical analysis it is convenient to use the mean flow $J = < v\rho> $  for a given car density. The relation between the vehicle density $\rho$ and the vehicle flow $J$ is called the fundamental diagram. The flow depends on the car density and deceleration probability $J(\rho, p)$. The famous fundamental diagram  for $v_{max}=5$ is shown in  Fig. \ref{fig:1}. In this case $\rho_{crit}=1\slash 6$ and $J(\rho_{crit}, p=0)$ is the maximal flow. When $p$ is increasing then the densities which allows the car flow to be maximal decreases. In the following we will call the density corresponding to the maximal traffic flow as the critical car density for a given $p$ and denote $\rho_{crit}(p)$.

\section{Simulation procedure}

We simulate a one lane traffic using the model described in the previous section. The one-dimensional lattice of the length $L=10\,000$ sites with the periodic boundary conditions forms the lane. If $N$ denotes the number of cars on the lane then the car density $\rho$ is defined as $\rho= N/L$. 
Our simulations start at random initial conditions; random localization of cars on a lane and random initial velocities. For each initial configuration, we update the individual vehicle velocity and position in accordance with the update rules. The basic simulations are performed for the car density  $\rho\in [0.01, 0.5]$ with the step $\Delta\rho=0.01$ and deceleration probability $p\in [0,1]$ with the step $\Delta p=0.05$. The results are obtained by averaging over $10000$ time steps after the first $10000$ ones left for the system stabilization. The procedure is then repeated for a number (100) of different realizations. The averages over different realizations provided the mean values of physical quantities which are presented in subsequent figures.

\section{Results} 

\begin{figure}
\centerline{\includegraphics[width=0.8\textwidth] {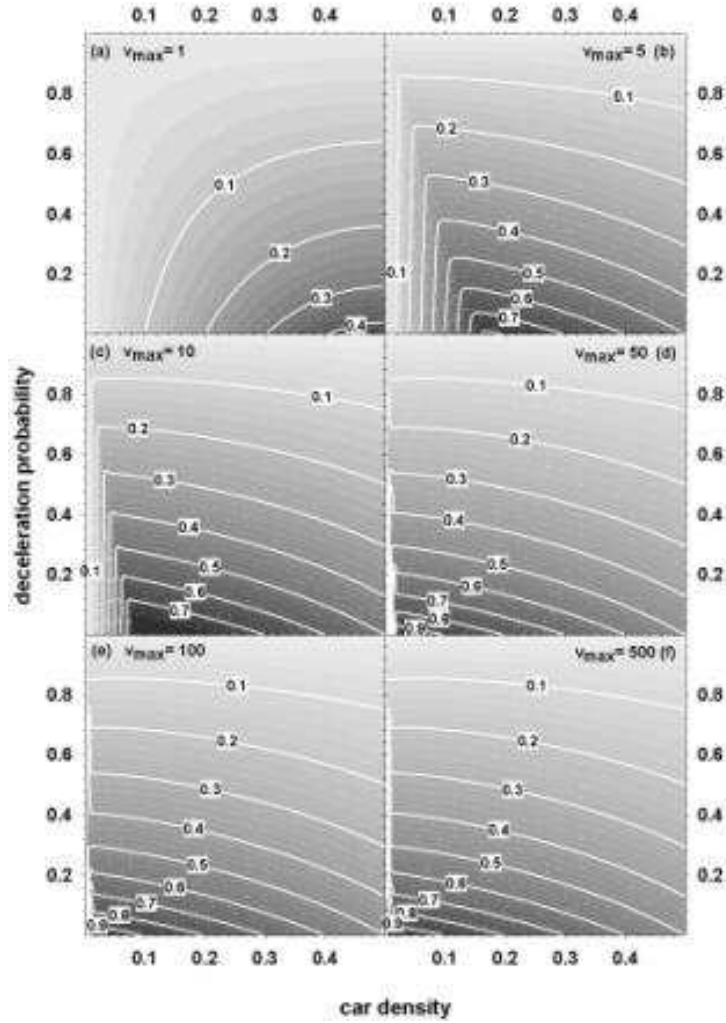}}
\caption{Traffic flow $J$ for different global velocity limit $v_{max}=1, 5, 10, 50, 100, 500$ -- the contour plots. Notice, changes in the free flow region and independence of the flow from the velocity limit  in case of  large  $v_{max}$ .  } 
\label{fig:2}        
\end{figure}

\subsection{Large velocity limit}

The limit $v_{max}=\infty$ has been introduced in \cite{SasvariKertesz} as equal to the length $L$ of the simulated finite system. Here we propose to investigate large velocities, however still finite to observe results of the velocity limit to properties of stationary traffic on a lane. In the series of contour plots in Fig. \ref{fig:2}  the fundamental diagrams are presented when the  velocity $v_{max}$ is  increased from $v_{\max}=1$ to $v_{\max}=500$.

If the velocity limit is low, e.g., $v_{\max}<10$,  then the fundamental diagram is a two part linear function, similar to that shown in Fig. \ref{fig:1}. The first part corresponds to the linear increase of the car flow in case of free flow regime. The second part is the linear decrease that is characteristic for the congested traffic. There is a narrow intersection interval of $\rho$ where the transition from one phase to the other takes place. When $v_{\max}$ grows then the transition point moves to lower densities. The congested traffic appears at all car densities with the exclusion of densities corresponding to almost empty lane, $\rho << 0.1$.

\begin{figure}[!ht]
\centerline{\includegraphics[width=0.6\textwidth] {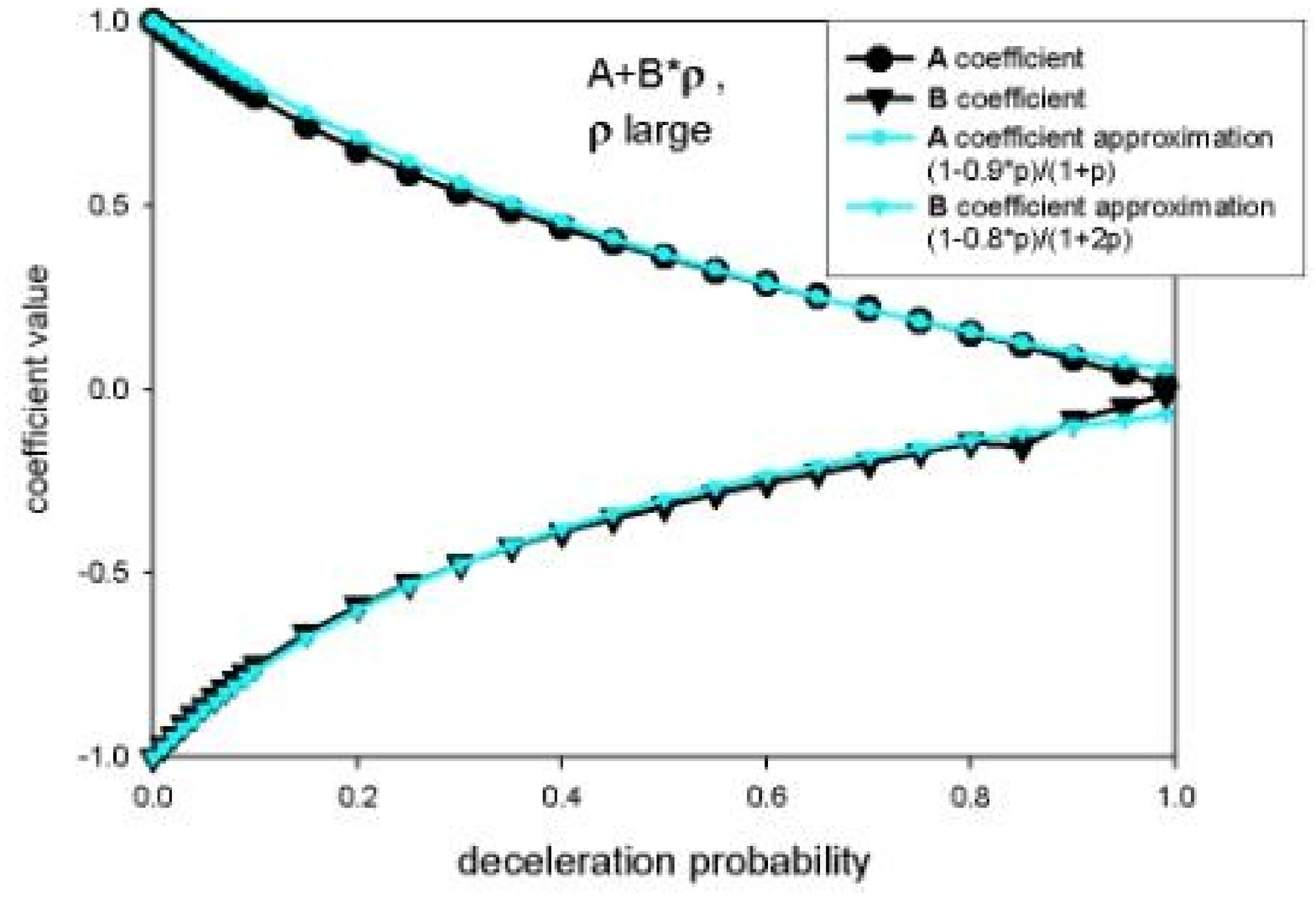}}
 \caption{Simulation and numerical estimates for the expected  stationary solution of the traffic flow $J= A(p)+B(p)\rho$, independently of $v_{max}$.  } 
 \label{fig:4} 
\end{figure}

It is noticeable that diagrams are almost indistinguishable when states of the congested traffic are considered  for $v_{\max}>50$. In Fig. \ref{fig:4} we present the numerical estimates for the linear decrease of the car flow $J(\rho, p)$ in the congested regime for $\rho >0.2$. The Pearson correlation coefficients for these  estimates are $r^2 >0.99 $. Therefore, the global dependence of $j$ on $p$ and $\rho$ can be found and our estimates lead to the following approximating formula:
\begin{equation}
J(\rho,p) = \frac{1-0.9p}{1+p} - \frac{1-0.8p}{1+2p} \rho
\label{flow}
\end{equation}
It is easy to notice that in deterministic traffic limit, $p=0$, we obtain the well known rigorous solution $J(\rho,0)= 1-\rho$.

\begin{figure}[!ht]
\centerline{\includegraphics[width=0.9\textwidth] {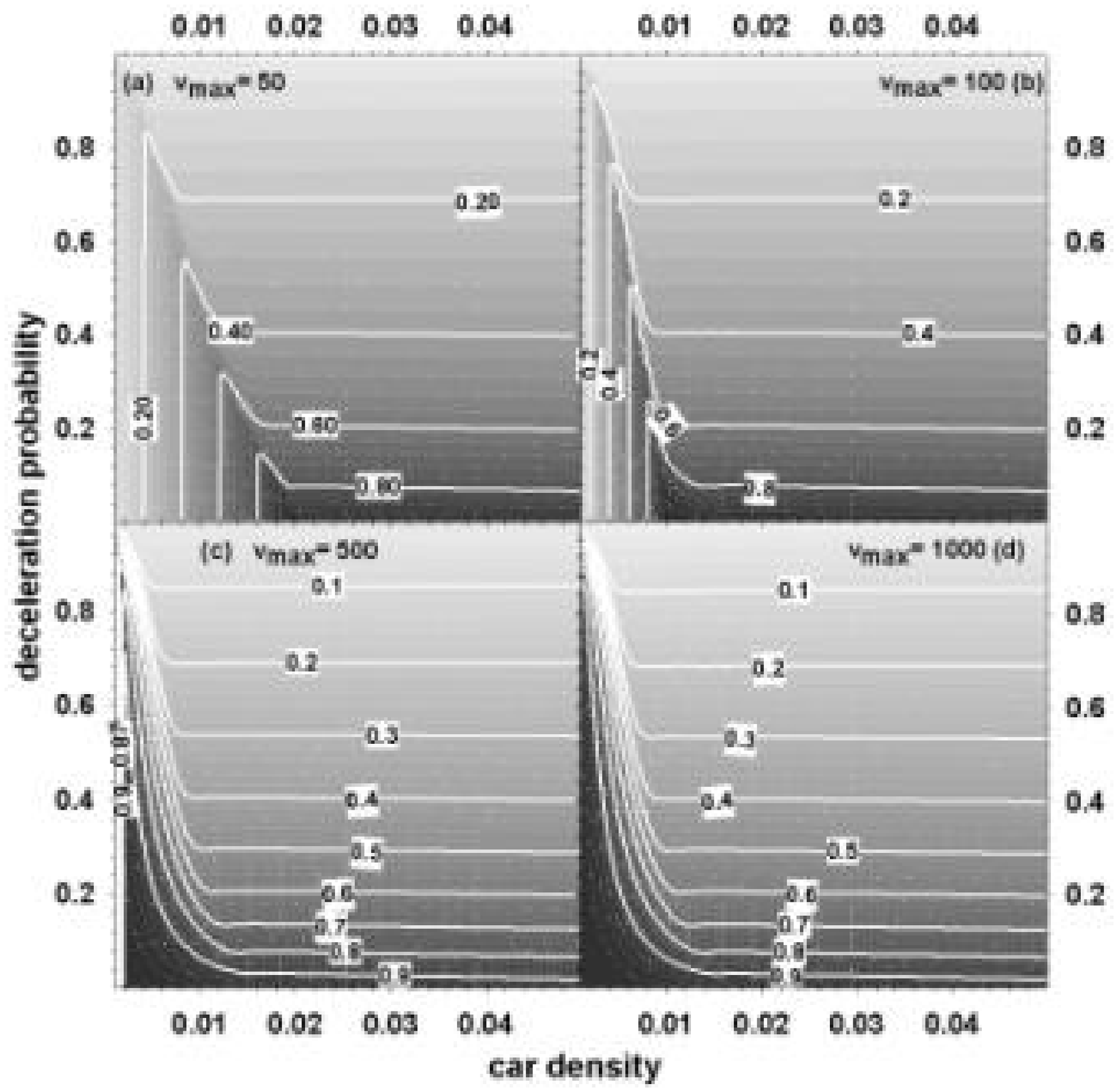}} 
\caption{Traffic flow $J$ for large global velocity limits $v_{max}=50, 100, 500, 1000$ and low car density  -- the contour plots.  The simulation step is $\Delta\rho=0.0005$. Notice high peaks in traffic flows appearing in  regions of  a transition from free flow traffic to congested traffic. } 
\label{fig:5} 
\end{figure}

\begin{figure}[!ht]
\includegraphics[width=0.49\textwidth] {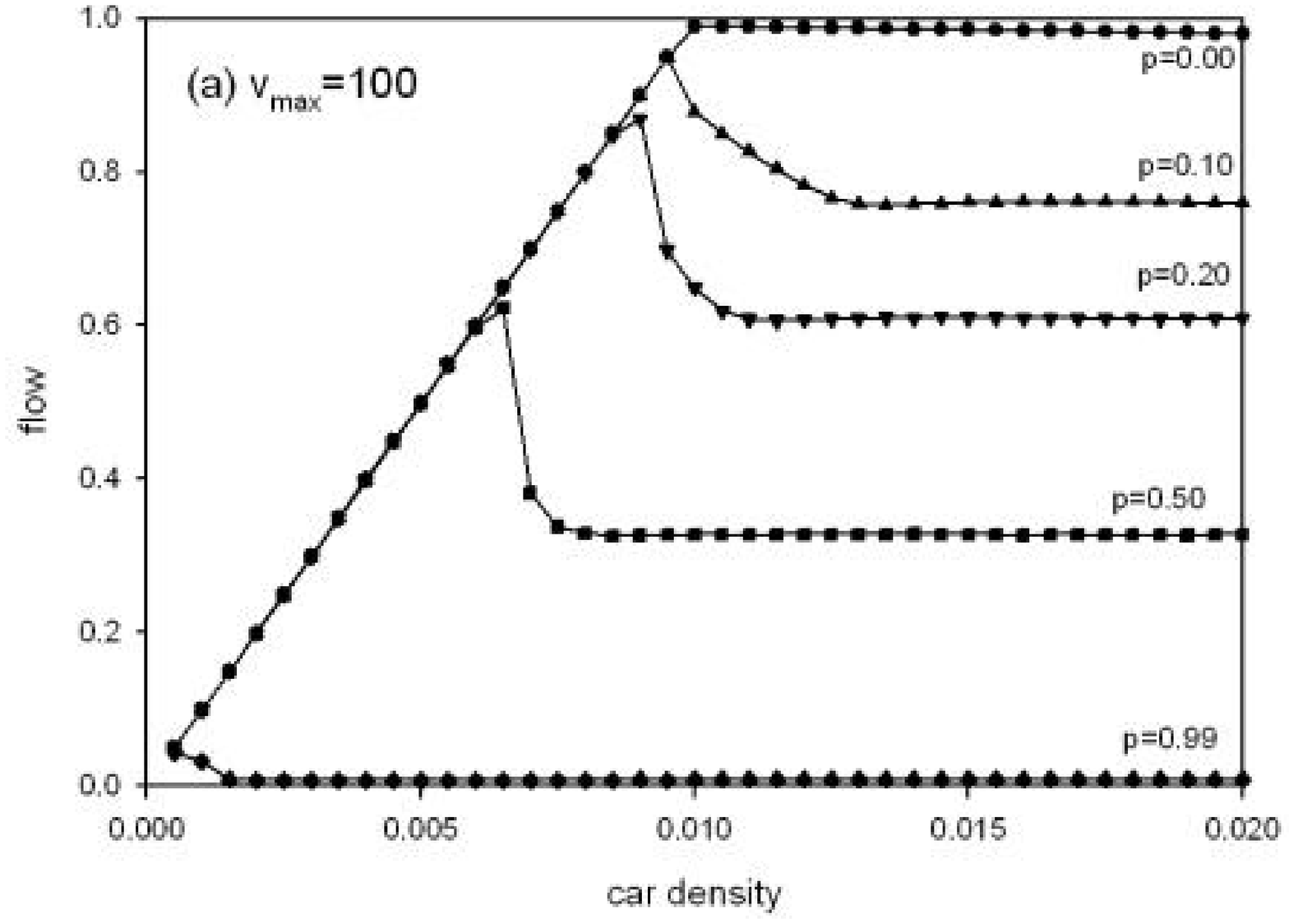}
\includegraphics[width=0.49\textwidth] {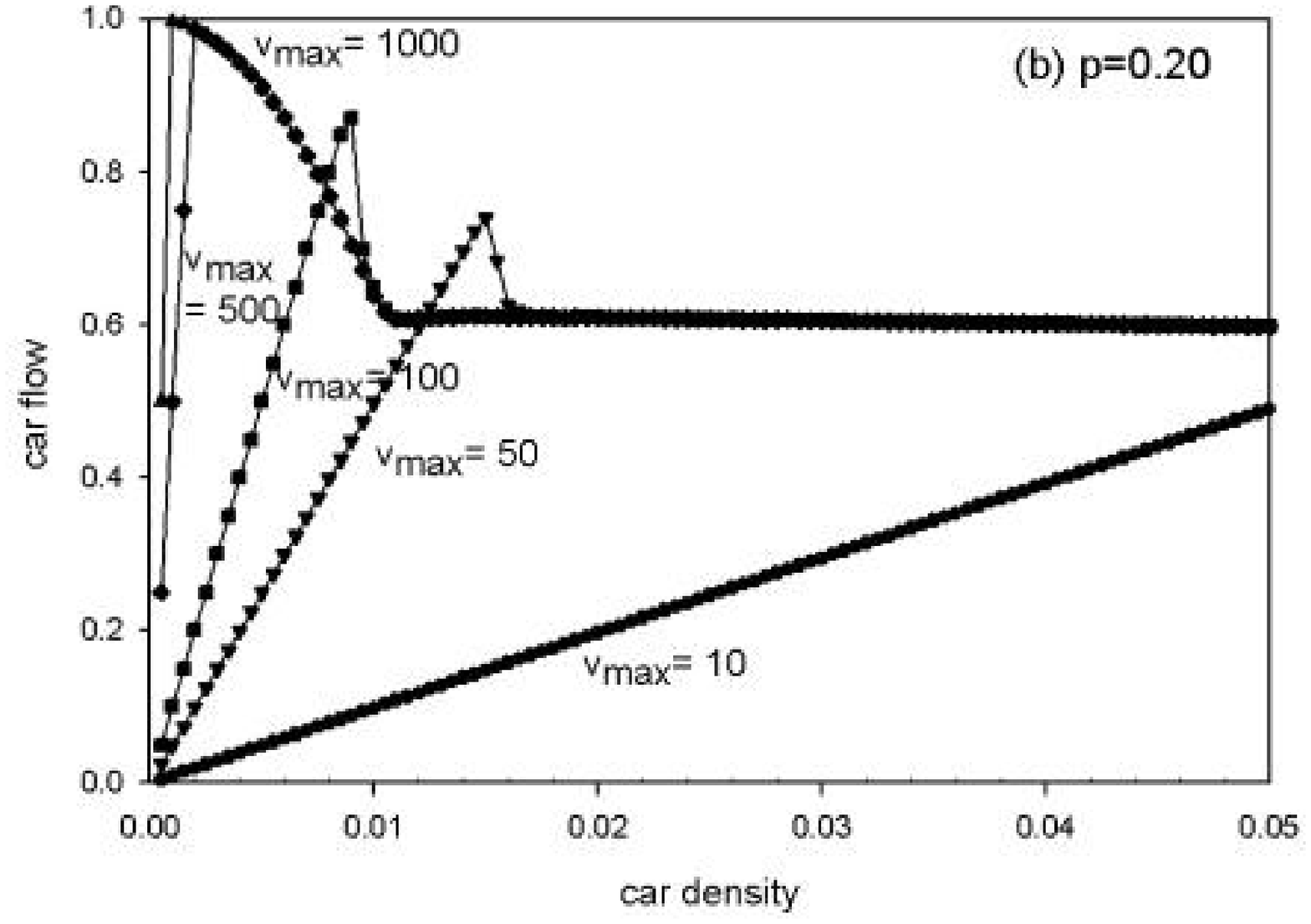}
\includegraphics[width=0.49\textwidth] {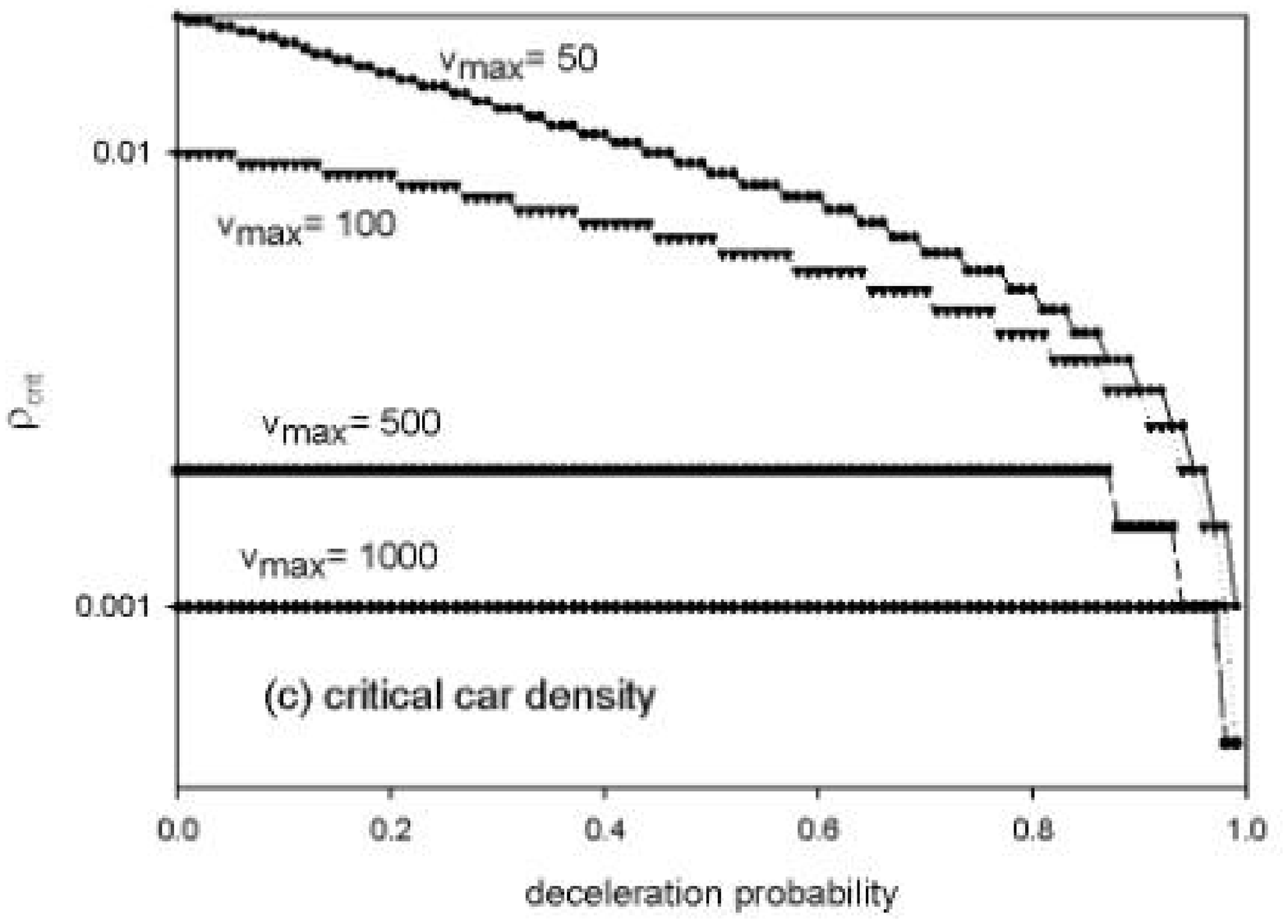}
\includegraphics[width=0.49\textwidth] {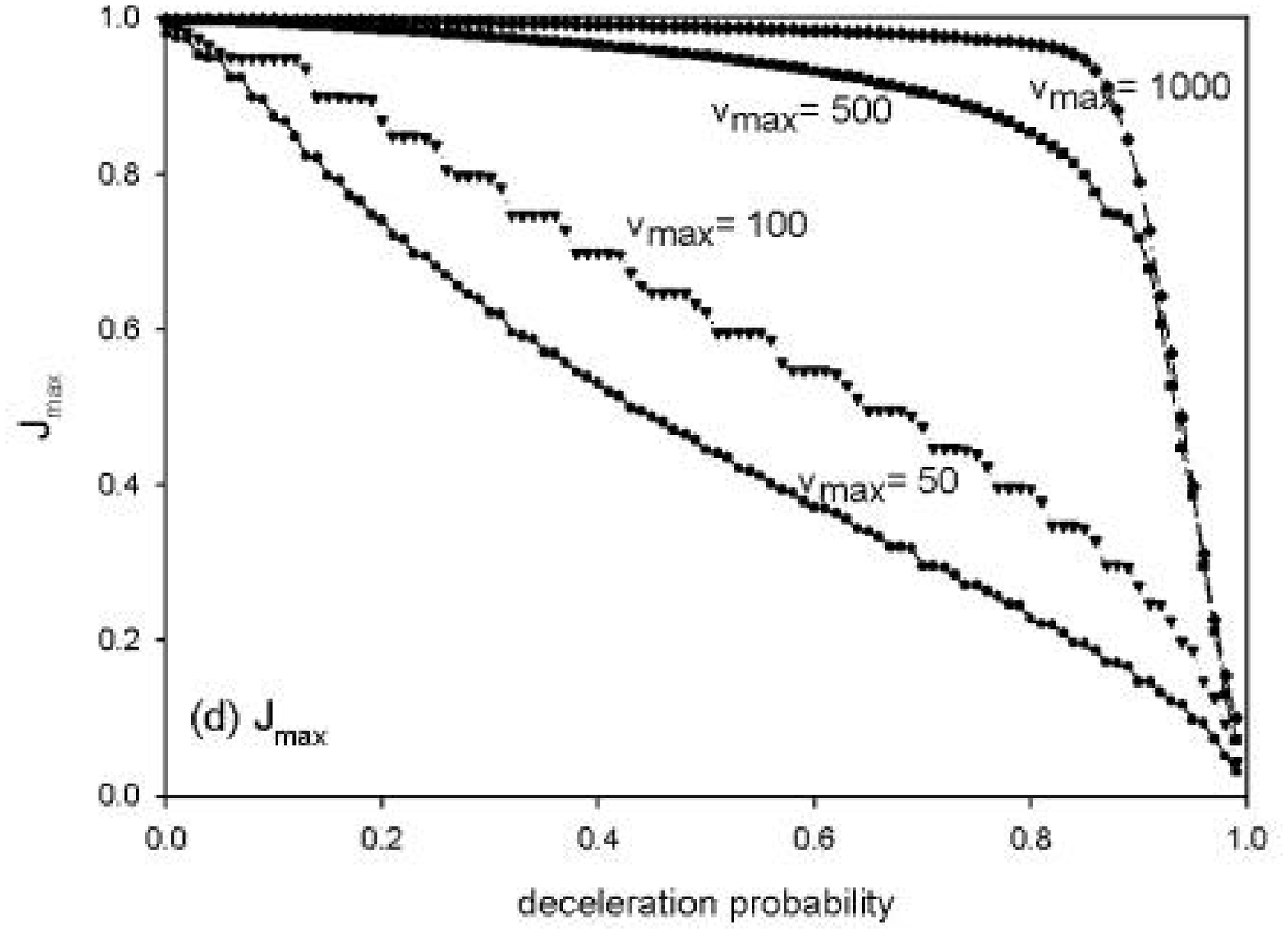} 
\caption{Extracted properties of the transition from the free flow traffic to the congested traffic in case of large maximal velocities. (a) The fundamental diagram of the NaSch model for the same speed limit $v_{max}=100$ and different deceleration probability. (b) The fundamental diagram of the NaSch model for different speed limits $v_{max}$ and  the same deceleration probability $=0.20$. (c) The critical car densities obtained for different $v_{max}$  (log-plot). (d) The dependence of the maximal flow on deceleration probability, for different $v_{max}$. } 
\label{fig:6} 
\end{figure}

The properties which are dependent on the velocity limit occur at lower densities. Therefore, we focus on the free flow and especially on localization of the maximum of the traffic flow. Figs \ref{fig:5} and \ref{fig:6} are to show different properties of the free flow and the transition. Figure \ref{fig:5} presents all data obtained by us while in Figs \ref{fig:6}(a)-(d) the extracted data is shown  in a way to visualize transition features.

The traffic flow increases rapidly when both the  car density  and velocity limit  $v_{max}$ grow. The critical car density $\rho_{crit}(p)$ and the maximal flow $J_{max}(\rho_{crit}(p),p)$ (see Figs \ref{fig:6}(c)(d)) are related to the end of the linear increase of the flow in the free flow regime.  

Comparing the flow diagrams around the critical points presented in Fig. \ref{fig:1} to corresponding ones shown in Figs \ref{fig:6}(a)(b)) one should notice that the value of the critical flow is significantly larger than it would be arisen from the congested state regime.  One can suspect that we observe the occurrence of the  free flow traffic in the congested state regime.
Hence, possible that  a kind of a regime with metastable states emerges for some interval of densities. 

Measurements on real traffic reveals that the traffic flow can exhibit metastability and hysteresis effects  but in the NaSch model in its simplest form these properties are not observed \cite{hysteresis}.  However, the metastable states and hysteresis are observed in the so-called Velocity-Dependent-Randomization model \cite{VDR}. Therefore it could be interesting to check the hypothesis whether starting the system  with all cars stopped in a traffic jam one obtains a stationary state different from a state reachable from a random initial condition. We performed the simulations for $v_{max}=100 $ and it appeared that the stationary flows were only slightly lower  than those presented in Fig. \ref{fig:6}(a). The largest differences were  observed in the interval  $0.3<p<  0.5$ and they were less than 0.02.   Thus, we cannot claim that in the NaSch model when the velocity limit is large  a region with a hysteresis memory emerges.

As it is expected, see Sec. 2 formula (\ref{max}), when $v_{max}$ increases the $\rho_{crit}(p)$ moves toward 0-density. Noticeable, that $p$ influence to $\rho_{crit}$ is weakened, see \ref{fig:6}(c) . For $v_{max}=1000$, $\rho_{crit}=0.001$  for almost all $p$. If $\rho > \rho_{crit}(p)$ then, as it is expected,  the monotonic and independent of $v_{max}$ decrease of maximal flow is observed, see Fig. \ref{fig:6}(b) . However,  the decay curve consists of the two easily recognizable parts: a nonlinear decrease (which can be approximated by some quadratic function with high accuracy) and linear decrease. The maximum of the  flow $J_{max}$ also becomes $p$ independent when the speed limit is sufficiently large, see \ref{fig:6}(d).

\subsection{Individual velocity limit} 
 
\begin{figure}[!ht]
\centerline{\includegraphics[width=0.8\textwidth]{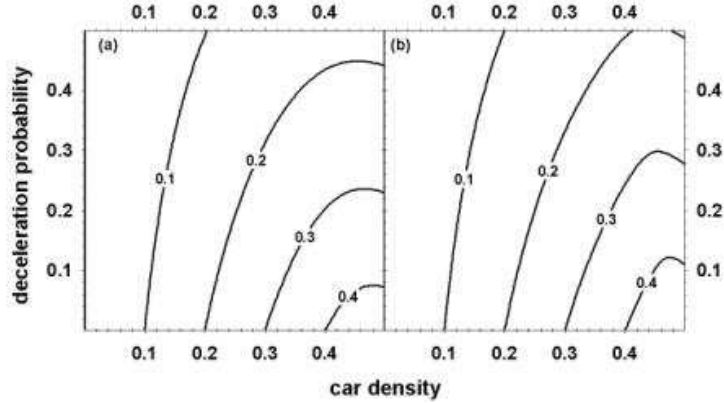}}
\caption{Traffic flow when each vehicle has its own maximal velocity $v_{max}(i)$ chosen at random  from $1,\dots,10$ (left panel) and from $1,\dots, 100$ (right panel). The slowest driving vehicle  has  the speed limit  exchanged with a new one.}
 \label{fig:7}       
\end{figure}

Let us assign the maximal velocity to each driver individually : $v_{max}(i), i=1,\dots, N$. These velocities are chosen at random from the interval $1, 2 , \dots, v_{max}$. 
Due to the hindrance effect of other cars $J(\rho)$ is determined by the slowest moving drivers. Hence the obtained fundamental diagram is almost the same as the fundamental one obtained in basic NaSch model with $v_{max}=1$.  It means that after stabilization steps all cars are moving in clusters  which are lead by cars with the speed limit equal to 1.  
Therefore, we search for the supplementary rules that cause a faster traffic movement.

let us start with the following rule:  at each time step one of the drivers --- that one who drives with the smallest actual velocity, is forced to revise his$\slash$her maximal velocity. Hence, before the acceleration substep (1), one car with the minimal actual velocity is found and then the new maximal velocity is assigned to it at random. Let us denote this rule as $(1,0)$ while the basic NaSch model is denoted $(0,0)$. The resulting flows plotted  for different $\rho$ and $p$ are  shown in  Fig. \ref{fig:7}. Their similarity to the shape of diagram with $v_{max}=1$ is  still noticeable.

To enhance  the effect of the $(1,0)$  rule, the following modifications are proposed (on the left, the notation of the new rules is placed):
\begin{itemize}
\item[$(2,0)$: ]the new velocity limit assigned to one of the slowest car must be greater than the car previous limit
\item[$(0,1)$: ]if a gap between two subsequent cars is equal to 0, then the speed limit of the hindered car is increased by 1. The new speed limit cannot excess the overall limit $v_{max}$.
\end{itemize}
The combination of the above notations means that the corresponding combination of the rules is applied simultaneously. For example, by $(1,1)$ we mean that one car with the minimal velocity is assigned a new velocity limit and all  cars which are separated from the following car by the zero gap have their speed limit increased by 1. 

\begin{figure}[!ht]
\includegraphics[width=0.48\textwidth]{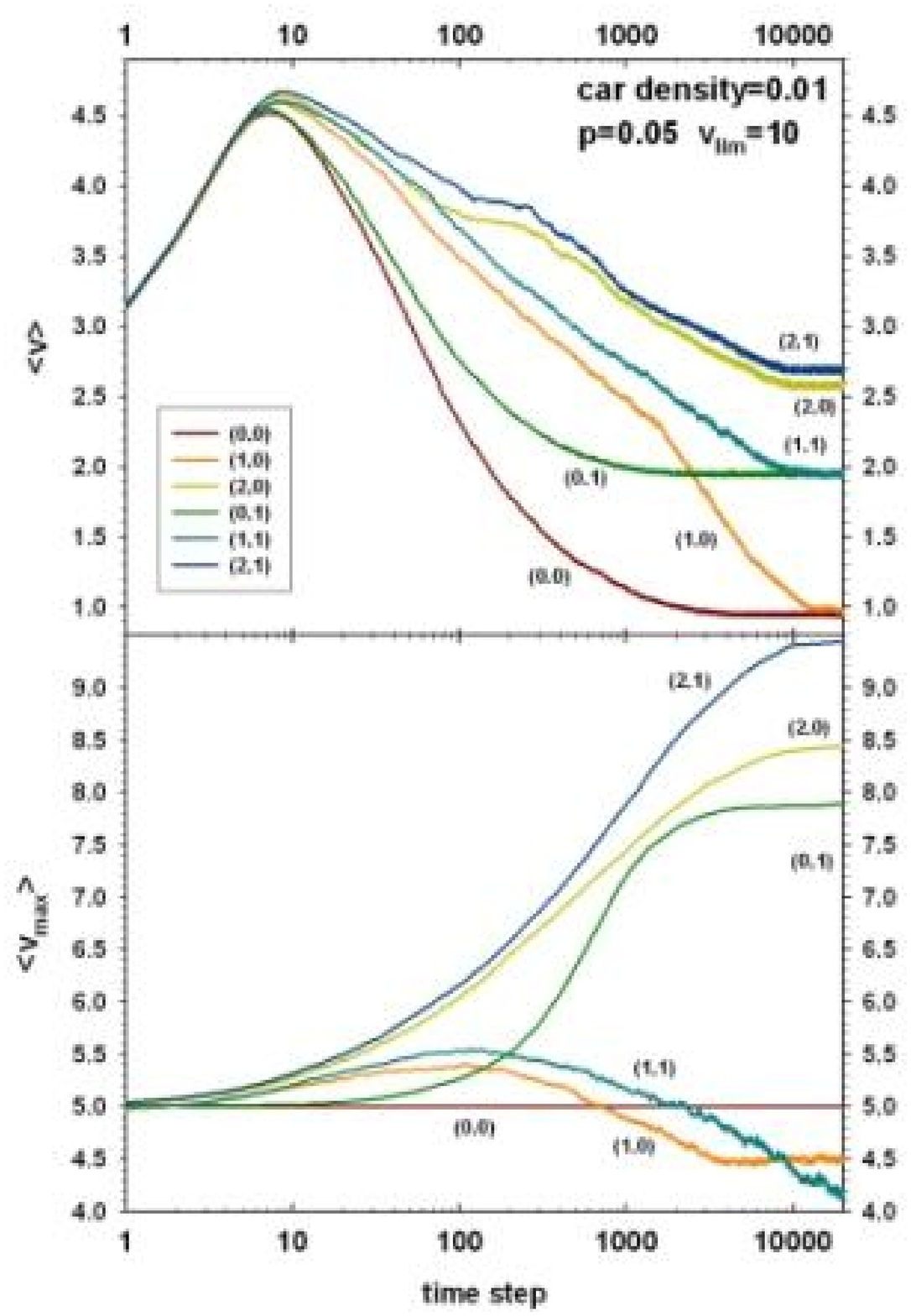} 
\includegraphics[width=0.48\textwidth]{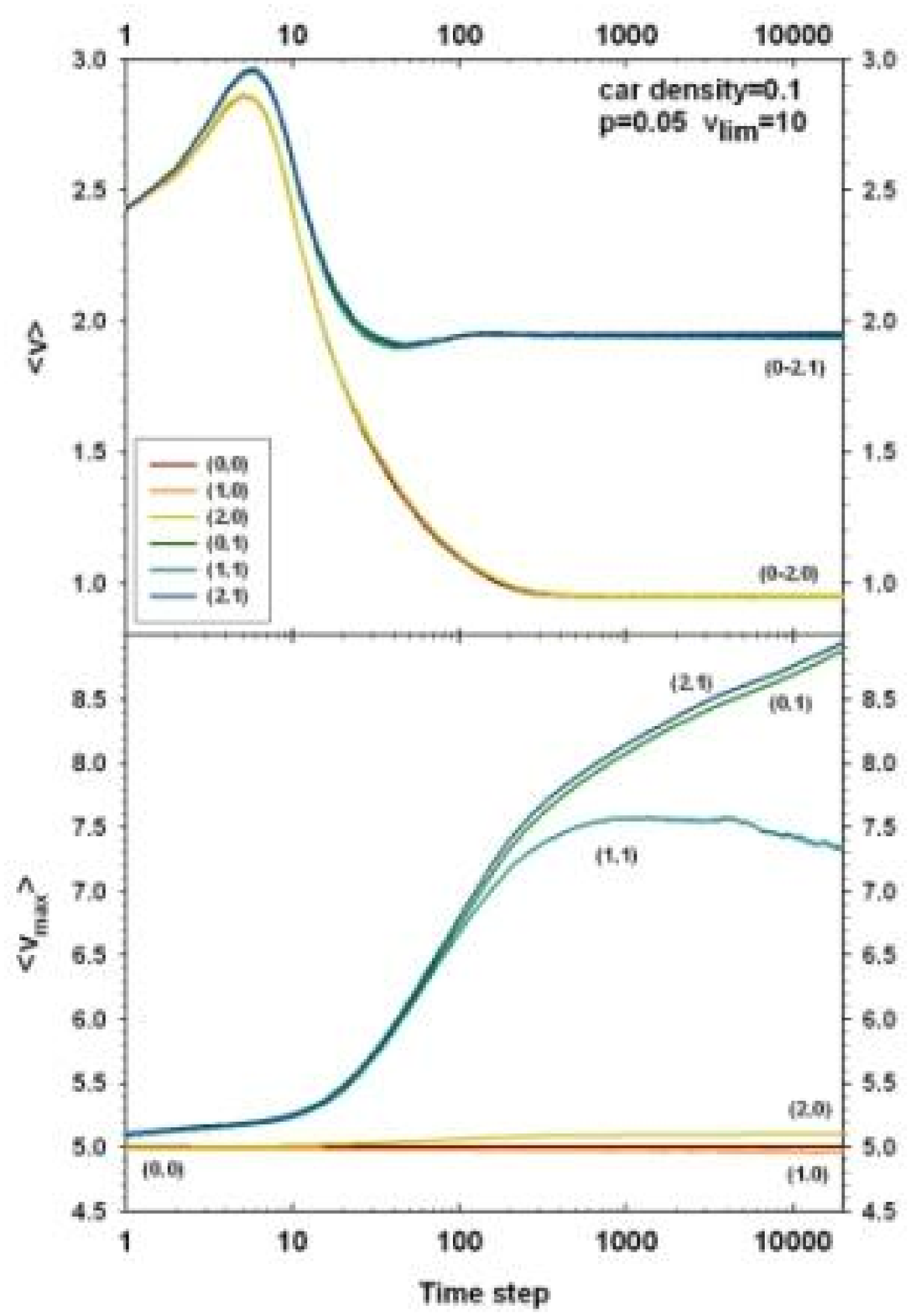}
\caption{Time development of the mean velocity $<v^{(t)})>$ and mean maximal velocity $<v_{max}^{(t)}>)$  in simulations when each vehicle has its own maximal velocity $v_{max}(i)$ chosen at random  from $1,\dots,10$. The labels at curves indicate at the supplementary rules applied to the basic NaSch model. (color plots on line) }
\label{fig:8}       
\end{figure}

\begin{figure}[!ht]
\includegraphics[width=0.98\textwidth]{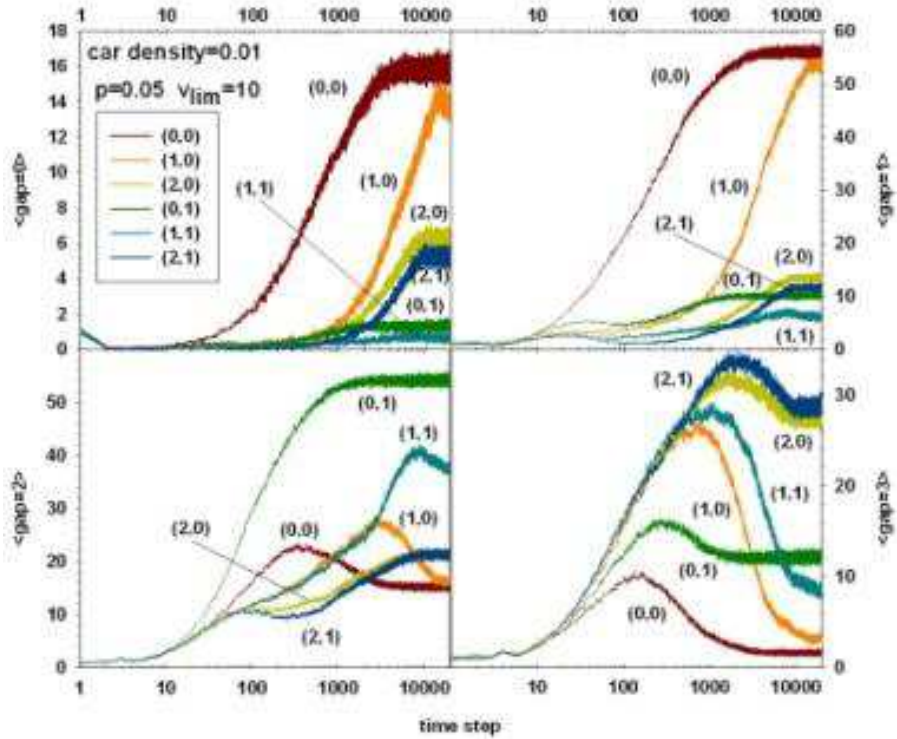} 
\caption{Time development of the small gaps --- distances in empty cells between subsequent vehicles. The labels at curves indicate at the supplementary rules applied to the NaSch model with individual speed limits. The vehicle density is very low $\rho=0.01 $ and the deceleration probability is low $p=0.05$. (color plots on line)}
\label{fig:9a}       
\end{figure}

\begin{figure}[!ht]
\includegraphics[width=0.98\textwidth]{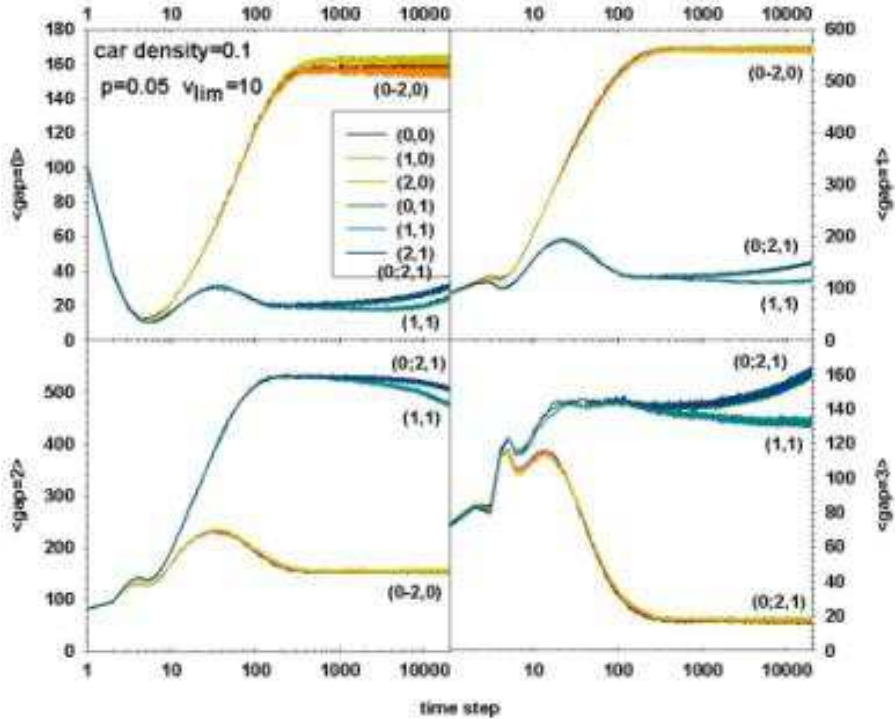} 
\caption{Time development of the small gaps --- distances in empty cellsbetween subsequent vehicles. The labels at curves indicate at the supplementary rules applied to the NaSch model with individual speed limits. The vehicle density is rather low, namely $\rho=0.1 $ but higher than in the density in the xperiment presented in Fig.\ref{fig:9a}. The deceleration probability is $p=0.05$. (color plots on line)}
\label{fig:9b}       
\end{figure}

Starting from random initial states with individual velocity limits taken  at random from the interval $1, 2 , \dots,10$, we observe the development of stationary states. In Fig. \ref{fig:8} the averages over a hundred realizations are presented. The  time evolution of the mean velocity and changes in the average maximal velocity are shown for the two low car densities: $\rho=0.01$ and $\rho=0.1$, and low deceleration $p=0.05$. The time axis is logarithmic to better visualize different features of subsequent time intervals.
In Figs. (\ref{fig:9a})(\ref{fig:9b}) we show the time development of small gaps $g_i= 0,1,2,3$, i.e. gaps responsible for the presence the following patterns $(0)$, $(e,1)$, $(e,e,2)$ and $(e,e,e,3)$, respectively, to have a deeper understanding of reasons for the global traffic properties.

When the car density is low, $\rho = 0.001$, Fig \ref{fig:8}(a), then in 10 first steps all cars accelerate to move with their individual maximal velocities. The mean velocity $<v>$ is only slightly lower than $<v_{max}>$. Only few small gaps are observed. Hence cars are separated suitable far away from each other on a lane.  But then cars with low maximal velocity limits cause hindrance to a free move of faster driving cars.  

In $(0,0)$ model, i.e. in the NaSch model, the number of stopped cars and cars moving with $<v>=1$  systematically grows while other velocities decay. If the slowest running car obtains a new maximal velocity at random, i.e.,  $(1,0)$ rule is applied, then  the increase of stopped cars and cars moving with $<v>=1$ is slowed down what results in the average velocity higher than 2. However, after many time steps, both rules lead to the similar stationary state: more than a half cars drive with velocity equal to 1, and there is a large number of  cars stopped. 

Surprisingly, that the mean maximal velocity $<v_{max}>$  settles at the value lower than the value obtained in case the  $(1,0)$ rule is applied. It happens because in our update configuration algorithm the slowest car is chosen as the most left of all cars moving with the slowest velocity. Hence, this car is often a last or close to the last car in a large traffic cluster. Usually, the tail cars are cars with the high speed limit. Therefore,  a random  change means here the exchange of the high speed limit to some another one, often lower. So that after many time steps we observe the decrease of the mean speed limit. 

If the slowest actually  driving car obtains a greater velocity limit, i.e. the rule $(2,0)$ is supplemented, then $<v_{max}>$ must be a non decreasing function and, indeed, this function is  growing fast  in the first 5000 time steps. Due to the low car density, the number of slowly  cars  is small and the mean velocity is high. After many steps almost $1\slash3$ of cars are separated by three empty cells: $(e,e,e,3)$,  what allows the global traffic to move at the mean velocity $<v> = 2.6$ -- the highest mean velocity  observed in a stationary traffic in our experiments. However, if the density of cars is larger, see Figs \ref{fig:8}(b) and \ref{fig:9b} then the differences in traffic properties  between the three rules considered  become negligible. The slowly moving cars drive slowly because of hindrance effect, not due to own speed limits. 

Let the NaSch model with random velocity limits be supplemented by $(0,1)$ rule. Then, it appears that the small number of stopped cars is permanent. The system stabilizes in a state where  more than a half cars are separated by 2 cells. This effects in the greater mean velocity than in $(0,0)$ model, namely $<v>=2$. Notice, that if between two subsequent cars the gap is 2 then two subsequent in time random brakes must happen to the first car to make the rule $(0,1)$ work, namely, to increase the first car maximal velocity. In case of $p=0.05$ it is a rare event. The system stabilizes quickly, in less than 1000 steps, for both car densities simulated.

 If together with accelerating those cars which cause stopping others, the slowest running car obtains a new random velocity, i.e.,  the $(1,0)$ rule is also supplemented,  then a traffic preserves the property of having only low number of stopped cars together with low number of cars separated by 1 empty cell. In the time interval of $100,\dots, 1000$ other properties of this traffic are similar to those observed when only $(1,0)$ rule is applied, though the mean velocity is higher. But in the stationary state the mean velocity becomes almost $2$.  In
case of larger density of cars, the stabilization is reached faster and  at approximately the same velocity. More than a half cars are separated by 2 two cells. The time development of systems governed by $(0,1)$  and $(2,1)$ is indistinguishable.
 
\section{Conclusions}
Despite its simplicity the NaSch model and its variants are able to reproduce many phenomena observed empirically. Although the same is true for other approaches mentioned here in the introduction, cellular automata model has the big advantage of being ideally suited for large-scale computer simulations. 

However, the results presented in this article in the series of figures, due to high accuracy and minuteness of detail, are demanding a large computer time. But their clear structure ensures us in their reliability.

In the following paper we basically investigated vanishing of the jamming transition when the velocity limit grows. Sasvari and Kertesz \cite{SasvariKertesz} suggested that the behavior of the NaSch model is governed by the two fixed points $p=0$ and $v_{max}=\infty$. Schadschneider \cite{Schadschneider} indicated that one should rather consider $p=0$ and $p=1$ as limiting fixed points in order to understand properties with finite velocity speed. Our results hint that for $v_{max}$ sufficiently large, the difference between the maximal possible flow, $J_{max} =1$ and $J(\rho_{crit}$ is negligible small for all $p\in[0,1)$, i.e. for all $p$ with the exception $p=1$ where all cars are stopped from the beginning and the evolution cannot influence the initial car patterns. On the other hand, for any $p\neq 0$ and $\rho\neq 0$ the intermediated patterns of (\ref{congested_patterns})-list are present. The free flow regime vanishes. Hence, the limit $v_{max}\rightarrow \infty$ excludes both deterministic evolutions: $p=0$ and $p=1$. Properties of the flow in the congested traffic phase are independent of speed limit and only deceleration probability influences the value of the mean flow.

By exchanging the lane velocity limit by the each driver velocity limit, we wanted to simulate the situation present on Polish roads. This situation results not only from  the popular among  drivers habit to break the speed limit but also from diversity of vehicles moving along the same lane: a horse wagon and a sport car. The simulations exemplify the dominant influence of the the slowest vehicle. However, the  insistence  of the car following the slowest vehicle , can improve, make higher, the velocity of the jammed traffic.

\section*{Acknowledgments}

We wish to acknowledge the support of Polish Ministry of Science and Information Technology -- project: PB/1472/PO3/2003/25.

\end{document}